\documentclass[preprint, aps, nofootinbib]{revtex4}
\usepackage{graphicx}
\usepackage{epsfig}
\usepackage{amsmath}
\usepackage{amsfonts}
\usepackage{amssymb}
\usepackage{color}%
\usepackage{dcolumn}
\usepackage{slashed}

\begin{document}
\title{Anisotropic Fermi surface from holography }
\author{Li Qing Fang$^{1,2}$}
\email{flqthunder@163.com}
\author{Xian-hui Ge$^{1,4}$}
\email{gexh@shu.edu.cn}
\author{Jian-Pin Wu$^{3,4}$}
\email{jianpinwu@gmail.com}
\author{Hong-Qiang Leng$^{1}$}
\email{lenghq88@shu.edu.cn}
\affiliation{${}^{1}$Institute of Theoretical Physics and Shanghai Key Laboratory of High Temperature Superconductors, Department of Physics, Shanghai University, Shanghai 200444, P.R. China\\
${}^{2}$School of Physics and Electronic Information, Shangrao Normal University, 334001  Shangrao, China\\
${}^{3}$Institute of Gravitation and Cosmology, Department of Physics, School of Mathematics and Physics, Bohai University, Jinzhou 121013, China\\
${}^{4}$State Key Laboratory of Theoretical Physics, Institute of Theoretical Physics, Chinese Academy of Sciences, Beijing 100190, China
}

\begin{abstract}
We investigate the probe holographic fermions by using an anisotropic charged black brane solution.  We derive the equation of motion of probe bulk fermions with one Fermi momentum along the anisotropic and one along the isotropic directions. We then numerically solve the equation and analyze the properties of Green function with these two momentums. We find in this case the shape of Fermi surface is anisotropic. However, for both Fermi momentums perpendicular to the anisotropic direction, the Fermi surface is isotropic. We verify that  our system obeys the recently conjectured bound for thermoelectric diffusion constants for the stable branch of the black brane solutions.
\end{abstract}

\maketitle




\section{Introduction}
Landau Fermi liquid theory is the standard model for theory of metals and it helps us  understand  almost all metals, such as semiconductors, superconductors and so on. Recently, this theory has been challenged by physics by the facts that lots of materials have been found which cannot be described by Landau Fermi liquid. For example, the Landau Fermi theory fails to describe electromagnetic properties of Weyl metal \cite{weyl1,weyl2}.  In recent years, using AdS/CFT correspondence \cite{adscft,gkp,w} to build a non-Fermi liquid theory has been widely studied in \cite{f1,f2,f3,f4,IL,JPW1,JPW2,FLQ1,FLQ2,XMK,KWP1,KWP2}. In ideal systems, the Fermi surface is isotropic, but in real materials, the Fermi surfaces are always anisotropic and inhomogeneous. For instance, the electric structure of superconducting cuprates, is highly anisotropic because of the atomic lattice effects.

In \cite{axionbackground1,axionbackground2}, one of us obtained a charged and spatially anisotropic black brane solution, dual to a spatially anisotropic $\mathcal{N}=4$ super Yang-Mills (SYM) theory at finite chemical potential and finite temperature. The anisotropy is introduced through deforming the SYM theory by a $\theta-$parameter of the form $\theta\propto x$, which acts as an isotropy-breaking source that forces the system into an anisotropic equilibrium state \cite{mateos1,mateos2}. Actually, the $\theta$-parameter is dual to the type IIB axion $\chi$ with the form $\chi = ax$, in which $a$ determines the level of the anisotropy. The prolate anisotropy corresponds to $a^2>0$ \cite{axionbackground1,axionbackground2}. The prolate black brane yields its very surprising property. That is to say, at a fixed temperature, there are two distinct branches of black brane solution: one branch with large, stable radii and the other branch with smaller but unstable radii. The thermodynamical phase structure of such a prolate black brane is similar to Schwarzschild-AdS black hole with spherical horizon.  Furthermore, the ratio of the shear viscosity to entropy density violates the  bound conjectured by Kovtun-Son-Starinets for the prolate black brane and the DC thermoelectric conductivities were obtained in \cite{glns}.

Considering the above facts, it is natural to ask what is the behavior of the holographic fermions in this background. The purpose of this paper is to investigate the properties of the probe holographic fermions in this anisotropic but homogenous background. Note that in our case, the linear axions do not lead to a periodic deformation of the boundary conformal field theory and thus cannot be considered as holographic lattice. Nevertheless, the axions do result in an anisotropic Fermi surface. It is worth noting that there also are some papers \cite{pomeranchuk,lattice} working on the anisotropic holographic fermions system. In \cite{pomeranchuk}, the authors  studied anisotropic Fermi surface by considering the Bramon-Grau-Pancheri (BGP) Lagrangian. But the gravity background is isotropic. In \cite{lattice}, the authors numerically constructed an anisotropic holographic lattice background by adding a neutral scalar field with the periodic boundary conditions along the spatial direction, and using this background to study the anisotropic Fermi surface. But  one advantage of our model is that the background anisotropic black brane is string-embedded and many of its properties remain
unknown to us.

Besides the effect of anisotropy, we wish to discuss the diffusion constant bound related to the Fermi velocity. Recently, motivating from holographic duality, the uncertainty principle and the measurements of diffusion in strongly interacting nonmetallic systems, Hartnoll  proposed a universal bound on the diffusion constant in an incoherent metal \cite{hartnoll2015}
\begin{equation}\label{bound}
D  \gtrsim\frac{ \hbar v^2_{F} }{k_B T},
\end{equation}
where $D $ are the diffusion constants, $v_F$ is the Fermi velocity. As the characteristic speed in a metal, $v_F$ plays a role analogous to the speed of light. Non-Fermi liquid, particularly bad metals cannot admit a quasiparticle
description because such quasiparticles would have a mean free path shorter than their Compton wavelength. In such incoherent metals, transport is controlled by the collective diffusion of energy and charge rather than by quasiparticle or momentum relaxation. The optical conductivity of such incoherent metals can  cross the  Mott-Ioffe-Regel (MIR) bound, so that it can be both weaker and stronger than the MIR bound. The bound presented in (\ref{bound}) was proposed to replace the MIR bound in bad metals and it can be simply derived from the Kovtun-Son-Starinets bound ${\eta}/{s}\geq C  {\hbar}/{k_B}$ and the relation ${\eta}/{s}=D T/c^2$ for vanishing chemical potential. Note that one need replace the speed of light $c$ with the Fermi velocity $v_F$ in metals. One point should be emphasized that  $C$ is simply a constant. In this paper, we will address this problem by utilizing the numerical results obtained for the probe Fermions.  We verify that  our system obeys the Hartnoll bound for thermoelectric diffusion constants for the stable branch of the black brane solution.


This paper is organized as follows:  we briefly review the anisotropic and charged black brane solution in Sec. \ref{sec2}. In Sec. \ref{sec3}, we give the equation of motion for the probe fermions with one momentum along the anisotropic direction and one momentum along the isotropic direction. We solve the Dirac equation numerically and study the properties of the Fermi surface in Sec. \ref{sec4}. In particular, we study the holographic fermions with momentums along the isotropic directions and reveal that the resulting Fermi surface is isotropic. The diffusivity bound in Eq.(\ref{bound})  will be examined by using the numerical results. The conclusion is presented in the last section.  For completeness in mathematics,  we  briefly discuss the Fermi surface structure of oblate anisotropy as a toy model in the appendix.

\section{The anisotropic charged black brane solution}\label{sec2}
The anisotropic charged black brane solution can be derived from the five-dimensional Einstein-Maxwell-Dilaton-Axion truncation
of gauge AdS supergravity with compactification of ten-dimensional type IIB supergravity
on $S^5$.  The nonlinear Kaluza-Klein reduction of type IIB supergravity to five dimension, leads to the presence of an Abelian field in the action. The time component of the Abelian field results in a nonzero
 chemical potential in the dual gauge theory.
 Different from the chargeless anisotropic black brane solution,  the introduction of the U(1) gauge field breaks the
SO(6) symmetry and thus leads to the excitations of the Kaluza-Klein modes.
\subsection{Background solution}
The effective action for the Einstein-Maxwell-Dilaton-Axion theory can be written as\cite{axionbackground1,axionbackground2}
\begin{equation}\label{action}
S=\frac{1}{2 \kappa^2} \int_{\mathcal{M}} \sqrt{-g} \Big(\mathcal{R}+12-\frac{1}{2}(\partial\phi)^{2}-\frac{1}{2}e^{2\phi}(\partial\chi)^{2}-\frac{1}{4} F_{MN}F^{MN}\Big)+\frac{1}{2 \kappa^2} \int_{\partial\mathcal{M}}\sqrt{-\gamma}2K,
\end{equation}
where $\kappa^2=8\pi G_5=4 \pi^2/N^2_c$, $L=1$, $F_{MN}=\partial_{M} A_{N}-\partial_{N} A_{M}$, and $\chi$ is an axion field.
The black brane solution  takes the following form
\begin{eqnarray}\label{metric}
&&ds^2=e^{-\frac{1}{2}\phi}r^2\Big(-\mathcal{F}\mathcal{B}dt^2+\mathcal{H}dx^2+dy^2+dz^2\Big)+\frac{e^{-\frac{1}{2}\phi}dr^2}{r^2\mathcal{F}}\\
&&\chi=ax,~~~A=A_t dt,~~~\phi=\phi(r),
\end{eqnarray}
where $A_t$ is dual to the chemical potential. The metric was already solved  numerically and analytically in \cite{axionbackground1,axionbackground2}.  While $a^2>0$ corresponds to the prolate anisotropy. We emphasize  that the linear axion field is not a dynamical field here and as can be seen from the action (\ref{action}) it plays the role of mass term of the dilatonic field.  Actually, the constant $a$ has dimensions of mass and is a measure of the anisotropy. From the five-dimensional theory viewpoint, the anisotropy can be interpreted as a nonzero number of dissolved D7-brane wrapped on $S^5$, extending along the $yz$-direction and distributed along the $x$-direction with density $n_{D7}$ \cite{mateos1}. Note that the form of the metric given here is slightly different from that of  \cite{axionbackground1,axionbackground2} and we choose $\chi=a x$ only for the convenience of computation.

In the following, we will mainly utilize the analytic black brane solution. The metric functions given in the $r$-coordinate can be solved by perturbing around the isotropic $\rm Resseiner-Nordstr\ddot{o}m$-AdS black brane in the small $a$  expansion \cite{axionbackground1,axionbackground2}
\begin{eqnarray}\label{metricfactor}
 &&\mathcal{F}=1-\bigg(\frac{r_H}{r}\bigg)^4+\bigg[\bigg(\frac{r_H}{r}\bigg)^6-\bigg(\frac{r_H}{r}\bigg)^4\bigg]q^2+a^2 \mathcal{F}_2(r)+\mathcal{O}(a^4),\nonumber\\[1.7mm]
 &&\mathcal{B}=1+a^2 \mathcal{B}_2(r)+\mathcal{O}(a^4),\nonumber\\[1.7mm]
 &&\mathcal{H}=e^{-\phi(r)}, {~~~\rm with ~~~}  \phi(r)=a^2 \phi_2(r)+\mathcal{O}(a^4),
\end{eqnarray}
 where
\begin{eqnarray}
 \mathcal{F}_2(r)&=&\frac{r_{H}^4}{24\sqrt{1+4q^2}}\bigg\{3(-\frac{4q^2}{r^6}+\frac{1}{r_{H}^6})
 \ln\left(\frac{(1+\sqrt{1+4q^2})r_{H}^2+2r^2}{(1-\sqrt{1+4q^2})r_{H}^2+2r^2}\right)\nonumber\\[0.7mm]
 &+&\frac{1}{r^2r_{H}^2}\Big[8\sqrt{1+4q^2}(-\frac{1}{r^2}+\frac{1}{r_{H}^2})+\frac{1}{r^2}\Big(3\ln\left(-2-2q^2+2\sqrt{1+4q^2}\right)\nonumber\\[0.7mm]
 &+&5(-2+q^2)\ln\left(-1+2q^2+\sqrt{1+4q^2}\right)-12q^2\ln\left(-2-2q^2+2\sqrt{1+4q^2}\right)\nonumber\\[0.7mm]
 &+&7(1+q^2)\Big(\ln\left(\frac{(-1+2q^2-\sqrt{1+4q^2})(2q^2u^2+(-1+\sqrt{1+4q^2}))}{r_{H}^2}\right)\nonumber\\[0.7mm]
 &-&\ln\left(\frac{2q^2}{r^2}-\frac{(1+\sqrt{1+4q^2})}{r_{H}^2}\right)\Big)\Big)\Big]\bigg\},\nonumber\\[1.7mm]
 \mathcal{B}_2(r)&=&\frac{1}{24}\left(\frac{10r^2}{q^2r_{H}^4-r^2r_{H}^2-r^4}+\frac{1}{r_{H}^2\sqrt{1+4q^2}}
 \ln\left(\frac{(1+\sqrt{1+4q^2})r_{H}^2+2r^2}{(1-\sqrt{1+4q^2})r_{H}^2+2r^2}\right)\right),\nonumber\\[1.7mm]
 \phi_2(r)&=&-\frac{1}{4r_{H}^2\sqrt{1+4q^2}}\ln\left(\frac{(1+\sqrt{1+4q^2})r_{H}^2+2r^2}{(1-\sqrt{1+4q^2})r_{H}^2+2r^2}\right),
\end{eqnarray}
and dimensionless charge $q=\frac{Q}{2\sqrt{3}r_H^3}$. The constant $Q$ is a dimensional charge which corresponds to the $U(1)$ gauge field. The gauge field $A_t$ is given by
\begin{eqnarray}
A_t&=&\sqrt{3}q r_{H}^3(\frac{1}{r_{H}^{2}}-\frac{1}{r^2})-\frac{5r_{H}}{24(\sqrt{4 q^2+1})} \Big\{\frac{\sqrt{3}q} {r^2}\ln\Big[\frac{(\sqrt{4q^2+1}+1)r^2-2q^2r_{H}^2}{(\sqrt{4 q^2+1}-1)r^2+2q^{2}r_{H}^2}\Big]\nonumber\\
&-&\frac{\sqrt{3}q}{r^2} \ln\Big[\frac{\sqrt{4 q^2+1}+1}{\sqrt{4 q^2+1}-1}\Big]+\frac{\sqrt{3}q}{r_{H}^2} \ln\Big[\frac{\sqrt{4 q^2+1}+1}{\sqrt{4 q^2+1}-1}\Big]\nonumber\\
&-&\frac{\sqrt{3}q}{r_{H}^2} \ln\Big[\frac{\sqrt{4 q^2+1}-2 q^2+1}{\sqrt{4 q^2+1}+2 q^2-1}\Big]\Big\}a^2.
\end{eqnarray}
The corresponding chemical potential is given by
\begin{equation}
\mu=\frac{q r_{H}}{8\sqrt{3}}\bigg( 24+\frac{5\ln\left(\frac{3-\sqrt{4 q^2+1}}{3+\sqrt{4 q^2+1}}\right)}{r^2_{H}\sqrt{4 q^2+1}}a^2\bigg).
\end{equation}
The charge density is given by $\rho=\frac{\sqrt{3}q r^3_{H}}{\kappa^2}$. In Taylor series expansion of $a$, the Hawking temperature and entropy density can be expressed as
\begin{equation}\label{temperature}
 T=\frac{(2-q^2)r_{H}}{2\pi}
 +\frac{\left(-4\sqrt{1+4q^2}+5(2+5q^2)\ln\big(\frac{3+\sqrt{1+4q^2}}{3-\sqrt{1+4q^2}}\big)\right)}{96\pi r_{H}\sqrt{1+4q^2}}a^2+\mathcal{O}(a^4).
\end{equation}
and
\begin{equation}
 s=\frac{N^{2}_{c}r_{H}^3}{2\pi}
 +\frac{5r_{H}N_{c}^2\ln\big(\frac{3+\sqrt{1+4q^2}}{3-\sqrt{1+4q^2}}\big)}{32\pi\sqrt{1+4q^2}}a^2+\mathcal{O}(a^4).
\end{equation}
\subsection{Thermodynamic  properties}
The anisotropic black brane yields very interesting thermodynamic  properties, as discussed  in \cite{axionbackground1,axionbackground2}. For the prolate anisotropy $a^2>0$, there are two branches of allowed black brane solutions to a fixed temperature, a branch with larger horizon radii and one with smaller.  The smaller branch of solution is unstable with negative specific heat. This situation is very similar to the case of Schwarzschild-AdS black holes with a spherical horizon.

\begin{figure}
\centering
\includegraphics[width=.5\textwidth]{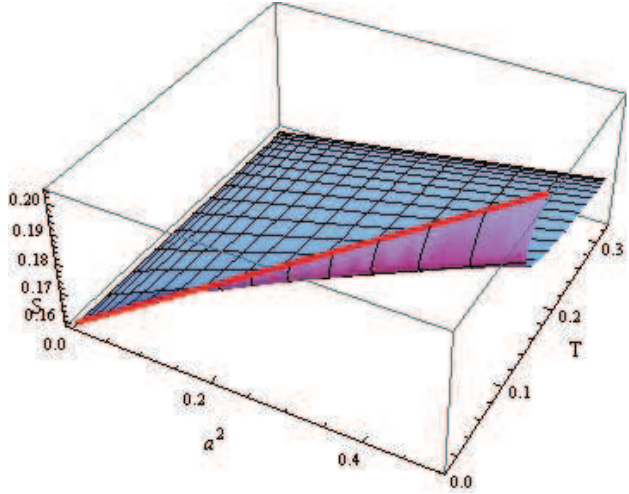}
\caption{The plot of the relation between temperature and entropy for a variational $a^2>0$. }
\label{aTS}
\end{figure}

Before discussing the property of holographic fermions, we should clarify the  parameters used in the numerical computation.
In order to work near the zero temperature, we will fix the dimensionless charge as $q=1.4$. After that, we vary the anisotropy parameter $a^2$.  Fig.\ref{aTS} shows how the entropy density varies as the temperature and $a^2$ changes. The parameter range ($q=1.4$ and $a^2$) is shown by the red line of the Fig. \ref{aTS}. For the prolate anisotropy $a^2>0$, there exists
 two branches as shown in  Fig.\ref{aTS}. We will work with the stable branch.  Note that the background thermodynamics is exactly the same as RN-AdS when $a^2=0$.
We will study the properties of holographic fermions for the cases $a^2=0$, $a^2>0$ in the following section, respectively. For completeness of our mathematical computations,  we  will also discuss the oblate anisotropy  case
with $a^2<0$ in Appendix \ref{oblate}.
\section{Dirac Equation}\label{sec3}
For the purpose of  studying the properties of the probe fermions on the dual boundary theory, we consider the bulk action for a probe Dirac fermion with the mass $m$, charge $q_f$. The
action of the bulk fermions is given by
\begin{equation}
S_{bulk}=\int
d^{5}x\sqrt{-g}i\bar{\psi}\Big(\Gamma^a D_{a}-
m\Big)\psi,
\end{equation}
where $\Gamma^a=(e_\mu)^a\Gamma^\mu$, the covariant derivative
$D_{a}=\partial_{a}+\frac{1}{4}(\omega_{\mu\nu})_a\Gamma^{\mu\nu}-iq_f A_a$.
The spin connection 1-forms
$(\omega_{\mu\nu})_{a}=(e_\mu)^b\nabla_a(e_\nu)_b$.
Here $(e_\mu)^a$ form a set of orthogonal normal vector bases.
From the above action, the Dirac equation can be derived as
\begin{equation}\label{diraceq}
\Gamma^a D_{a}\psi-m\psi=0.
\end{equation}
Then, we make a Fourier transformation $\psi=(-g g^{rr})^{-\frac{1}{4}}e^{-i\omega t+ik_{x}x+ik_{y}y}\tilde{\phi}$, and use the following gamma matrices basis
\begin{eqnarray}
\Gamma^r&=&\left(
\begin{array}{cc}
-\sigma_{3}\ & 0 \\
0 & -\sigma_3\ \\
\end{array}\right),
\ \ \Gamma^t=\left(
\begin{array}{cc}
 i\sigma_1\ & 0 \\
 0 & i\sigma_1\ \\
\end{array}\right),\nonumber\\
\ \ \Gamma^x&=&\left(
 \begin{array}{cc}
 -\sigma_2\ & 0 \\
  0 & \sigma_2\ \\
\end{array} \right),~~
\ \ \Gamma^y=\left(
 \begin{array}{cc}
 0 & \sigma_2\ \\
  \sigma_2\ & 0\\
\end{array}\right).
\end{eqnarray}
Note that we choose $k_x$ along the anisotropic direction and $k_y$ along the isotropic direction.
We set $\tilde{\phi}=\left(
 \begin{array}{c}\tilde{\phi}_1 \\ \tilde{\phi}_2\\ \end{array} \right)$. The Dirac equation becomes two coupled equations
\begin{eqnarray}\label{phi12eom}
\sqrt{g^{rr}}\partial_{r}\tilde{\phi}_{1}+m\sigma_{3}\tilde{\phi}_{1}=\sqrt{g^{tt}}(\omega+q_f A_t)i\sigma_2
\tilde{\phi}_{1}-\sqrt{g^{xx}}k_{x}\sigma_{1}\tilde{\phi}_{1}+\sqrt{g^{yy}}k_{y}\sigma_{1}\tilde{\phi}_{2}\nonumber\\[1.2mm]
\sqrt{g^{rr}}\partial_{r}\tilde{\phi}_{2}+m\sigma_{3}\tilde{\phi}_{2}=\sqrt{g^{tt}}(\omega+q_f A_t)i\sigma_2
\tilde{\phi}_{2}+\sqrt{g^{xx}}k_{x}\sigma_{1}\tilde{\phi}_{2}+\sqrt{g^{yy}}k_{y}\sigma_{1}\tilde{\phi}_{1}.
\end{eqnarray}
In order to decouple the equation of motion, we assume $\tilde{\phi}_I=\left(\begin{array}{c}y_I \\ z_I\\\end{array} \right)$,
with $I=1,2$.  The equation of motion (\ref{phi12eom}) yields
\begin{eqnarray}\label{yzeom}
\sqrt{g^{rr}}\partial_{r}y_{1}+my_{1}&=&\sqrt{g^{tt}}(\omega+q_f A_t)
z_{1}-\sqrt{g^{xx}}k_{x}z_{1}+\sqrt{g^{yy}}k_{y}z_{2},\nonumber\\[1.2mm]
\sqrt{g^{rr}}\partial_{r}z_{1}-mz_{1}&=&-\sqrt{g^{tt}}(\omega+q_f A_t)
y_{1}-\sqrt{g^{xx}}k_{x}y_{1}+\sqrt{g^{yy}}k_{y}y_{2},\nonumber\\[1.2mm]
\sqrt{g^{rr}}\partial_{r}y_{2}+my_{2}&=&\sqrt{g^{tt}}(\omega+q_f A_t)
z_{2}+\sqrt{g^{xx}}k_{x}z_{2}+\sqrt{g^{yy}}k_{y}z_{1},\nonumber\\[1.2mm]
\sqrt{g^{rr}}\partial_{r}z_{2}-mz_{2}&=&-\sqrt{g^{tt}}(\omega+q_f A_t)
y_{2}+\sqrt{g^{xx}}k_{x}y_{2}+\sqrt{g^{yy}}k_{y}y_{1}.
\end{eqnarray}
The ingoing boundary condition for $\phi_{I}$ at the event horizon can be imposed as
\begin{equation}\label{Ingoing}
\tilde{\phi}_{I}\propto \left(\begin{array}{c} i \\1 \\
\end{array}\right)e^{-i\omega r_{*}},
\end{equation}
with $r_{*}=\int\frac{dr}{r^2 \mathcal{F}\sqrt{B}}$.
Near the AdS boundary, the solution of the Dirac equation (\ref{phi12eom}) can be written as
\begin{eqnarray}\label{bdysol}
\tilde{\phi}_{1}^{\rm{I}}\overset{r\rightarrow\infty}{\approx}\left(\begin{array}{c}c_{1}^{\rm{I}}r^{-m} \\ d_{1}^{\rm{I}}r^{m}\\\end{array}\right),~~~~~~
\tilde{\phi}_{1}^{\rm{II}}\overset{r\rightarrow\infty}{\approx}\left(\begin{array}{c}c_{1}^{\rm{II}}r^{-m} \\ d_{1}^{\rm{II}}r^{m}\\\end{array}\right),\nonumber\\
\tilde{\phi}_{2}^{\rm{I}}\overset{r\rightarrow\infty}{\approx}\left(\begin{array}{c}c_{2}^{\rm{I}}r^{-m} \\ d_{2}^{\rm{I}}r^{m}\\\end{array}\right),~~~~~~
\tilde{\phi}_{2}^{\rm{II}}\overset{r\rightarrow\infty}{\approx}\left(\begin{array}{c}c_{2}^{\rm{II}}r^{-m} \\ d_{2}^{\rm{II}}r^{m}\\\end{array}\right),
\end{eqnarray}
where $\rm{I}, \rm{II}$ correspond to two independent ingoing boundary conditions.
From the holographic dictionary, the retarded Green function is given by
\begin{equation}
G=\mathbb{C}\mathbb{D}^{-1},
\end{equation}
where we have defined
\begin{equation}
G\equiv \left(\begin{array}{cc} G_{11}\ & G_{12} \\ G_{21} & G_{22} \ \\ \end{array} \right),~~~
\mathbb{C}\equiv \left(\begin{array}{cc} c_{1}^{\rm{I}}\ & c_{1}^{\rm{II}} \\ c_{2}^{\rm{I}} & c_{2}^{\rm{II}}\ \\ \end{array} \right),~~~
\mathbb{D}\equiv \left(\begin{array}{cc} d_{1}^{\rm{I}}\ & d_{1}^{\rm{II}} \\ d_{2}^{\rm{I}} & d_{2}^{\rm{II}}\ \\ \end{array} \right).
\end{equation}
For numerical convenience, we can define the following matrices
\begin{equation}
Y\equiv \left(\begin{array}{cc} y_{1}^{\rm{I}}\ & y_{1}^{\rm{II}} \\ y_{2}^{\rm{I}} & y_{2}^{\rm{II}}\ \\ \end{array} \right),~~~
Z\equiv \left(\begin{array}{cc} z_{1}^{\rm{I}}\ & z_{1}^{\rm{II}} \\ z_{2}^{\rm{I}} & z_{2}^{\rm{II}}\ \\ \end{array} \right),~~~
\tilde{G}\equiv YZ^{-1}.
\end{equation}
And then one can obtain the evolution equation as follows
\begin{eqnarray}\label{anieom}
\sqrt{g^{rr}}\partial_{r}\tilde{G}_{11}&+&2m\tilde{G}_{11}-\sqrt{g^{tt}}(\omega+q_f A_t)(\tilde{G}_{11}^2+\tilde{G}_{12}\tilde{G}_{21}+1)\nonumber\\
&-&\sqrt{g^{xx}}k_{x}(\tilde{G}_{11}^2-\tilde{G}_{12}\tilde{G}_{21}-1)+\sqrt{g^{yy}}k_{y}\tilde{G}_{11}(\tilde{G}_{12}+\tilde{G}_{21})=0,\nonumber\\[1.2mm]
\sqrt{g^{rr}}\partial_{r}\tilde{G}_{22}&+&2m\tilde{G}_{22}-\sqrt{g^{tt}}(\omega+q_f A_t)(\tilde{G}_{22}^2+\tilde{G}_{12}\tilde{G}_{21}+1)\nonumber\\
&+&\sqrt{g^{xx}}k_{x}(\tilde{G}_{22}^2-\tilde{G}_{12}\tilde{G}_{21}-1)+\sqrt{g^{yy}}k_{y}\tilde{G}_{22}(\tilde{G}_{12}+\tilde{G}_{21})=0,\nonumber\\[1.2mm]
\sqrt{g^{rr}}\partial_{r}\tilde{G}_{12}&+&2m\tilde{G}_{12}-\sqrt{g^{tt}}(\omega+q_f A_t)\tilde{G}_{12}(\tilde{G}_{11}+\tilde{G}_{22})\nonumber\\
&+&\sqrt{g^{xx}}k_{x}\tilde{G}_{12}(\tilde{G}_{22}-\tilde{G}_{11})-\sqrt{g^{yy}}k_{y}(1-\tilde{G}_{12}^2-\tilde{G}_{11}\tilde{G}_{22})=0,\nonumber\\[1.2mm]
\sqrt{g^{rr}}\partial_{r}\tilde{G}_{21}&+&2m\tilde{G}_{21}-\sqrt{g^{tt}}(\omega+q_f A_t)\tilde{G}_{21}(\tilde{G}_{11}+\tilde{G}_{22})+\sqrt{g^{xx}}k_{x}\tilde{G}_{21}(\tilde{G}_{22}-\tilde{G}_{11})\nonumber\\
&-&\sqrt{g^{yy}}k_{y}(1-\tilde{G}_{21}^2-\tilde{G}_{11}\tilde{G}_{22})=0.
\end{eqnarray}
In terms of Eq.(\ref{Ingoing}), we can easily find the boundary condition for the above evolution equation at the horizon
\begin{eqnarray}\label{anibdycond}
\tilde{G}\overset{r\rightarrow r_H}{\approx}\left(\begin{array}{cc} i\ & 0 \\ 0 & i\ \\ \end{array} \right),
\end{eqnarray}
After solving the evolution equations (\ref{anieom}), we can read off the boundary Green's function as
\begin{eqnarray}\label{anigf}
G=\lim_{r\rightarrow\infty}r^{2m}\tilde{G}.
\end{eqnarray}

\section{The properties of anisotropic Fermi surface}\label{sec4}
 In this section, we mainly focus on the property of the spectral function $A(\omega,k_x,k_y)=Im[G_{11}+G_{22}]$. In the discussion below, we will study the case of $m=0$, $q_f=1$ and $r_H=1$ for simplicity but without loss of generality. Before proceeding, we would like to give a brief comment on the parameter $a$ and the dimensionless charge $q$. Since the contribution of $a$ to the temperature is small, the temperature of the system will be fixed in a small range to a fixed $q$.  Thus, we choose $q=1.4$ to make sure that the temperature of the system is finite and close to zero. In addition, we can conclude that $G_{11}(\omega, -k_x, k_y)=G_{22}(\omega, k_x, k_y)$ and $G_{12}(\omega, k_x, k_y)=G_{21}(\omega, k_x, k_y)$ from Eq. (\ref{anieom}). These symmetries can also be seen in our following numerical results.

\subsection{The shape of Fermi surface}
Now, we investigate the Fermi momentum along $x-$ and $y-$ axes simultaneously.
With this in mind, we solve the equations of motion (\ref{anieom}) numerically.

\begin{figure}
\centering
\includegraphics[width=.39\textwidth]{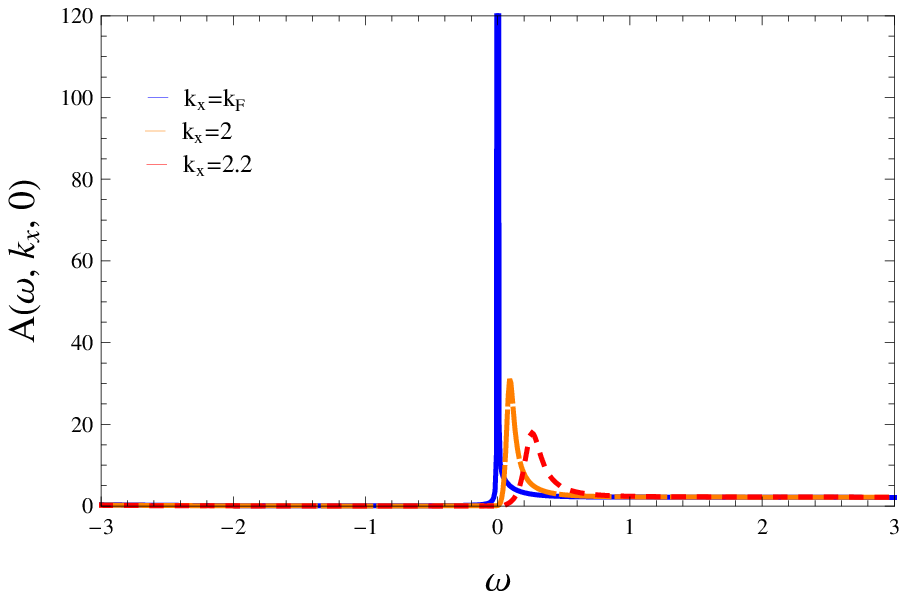}\hspace{0.6cm}
\includegraphics[width=.36\textwidth]{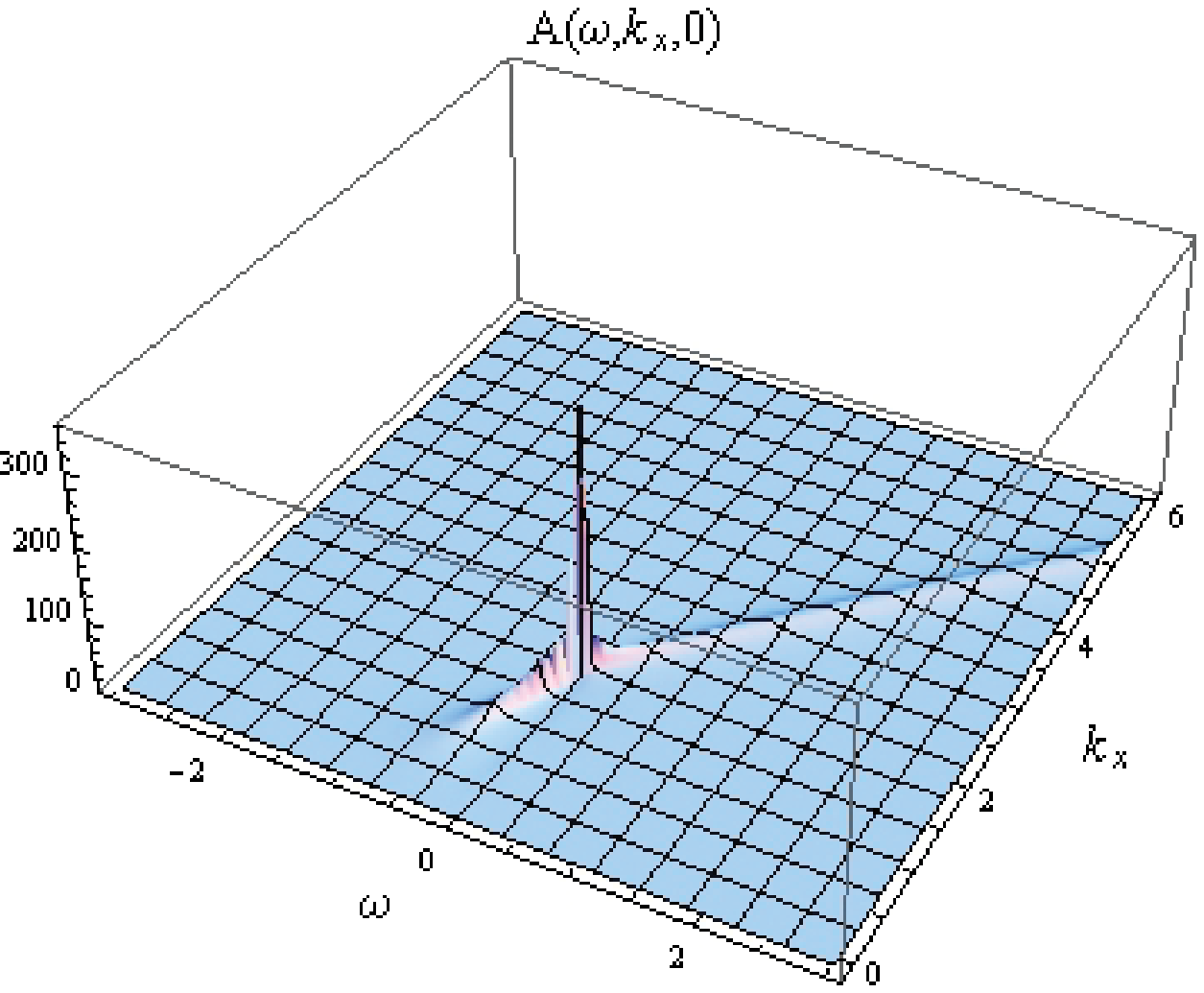}
 \caption{Left: Spectral function $A(\omega, k_x, 0)$ at $k_x=k_{F}$(blue solid), $k_x=2$(orange dashed) and $k_x=2.2$(red dotted) for $a=0$. Right: The 3D plot of $A(\omega, k_x, 0)$ for $a=0$.}
 \label{a0}
\end{figure}
First, we  look for the Fermi momentum along the $x-$direction (i.e. $k_y=0$) for $a^2=0$.
For the imaginary part of retard Green functions, a peak appear in the region $\omega>0$ which has a broad maximum (see the left plot of Fig.\ref{a0}). When $k_x=2$, this peak is sharper than $k_x=2.2$. As the value of momentum approaches $1.84318392$, a sharp quasiparticle like peak is generated near $\omega=0$, so that its height goes to infinity and width close to zero. By studying the spectrum function $A(\omega, k_x, 0)$ for a given $\omega=10^{-9}$, we can determine Fermi momentum $k_{F}=1.84318392$ along the $x-$ direction for $a^2=0$ (see the right plot of Fig.\ref{a0}). That is to say, when $a=0$, the background is the same as that of five-dimensional RN-AdS black hole. The above result agrees with that of \cite{f2,f3}, although the form of Green function is different to (\ref{anigf}). This is direct evidence that our equation of motion (\ref{anieom}) is correct.

Second, we will study the shape of the Fermi surface. When $a^2=0$, the background reduces to the metric of  RN-AdS black hole. We find that the shape of the Fermi surface is  isotropic for $a^2=0$.  In other words, the Fermi momentums are equal on each direction of the $k_x-k_y$ plane. We can fit our numerical result by a function as follows\cite{lattice}:
\begin{equation}
\frac{k_x^2}{c_x^2}+\frac{k_y^2}{c_y^2}=1.
\end{equation}
To be more explicit, we can also introduce two quantities that are the difference between the $k_x$ axis and $k_y$ axis $d=c_x-c_y$ and the flattening factor $f=\frac{c_x-c_y}{c_x}$. From the above result, we see that the shape of the Fermi surface for $a^2=0$ case is a sphere. It means $c_x=c_y=1.84318392$ and $d=f=0$.

\begin{figure}
\centering
\includegraphics[width=.32\textwidth]{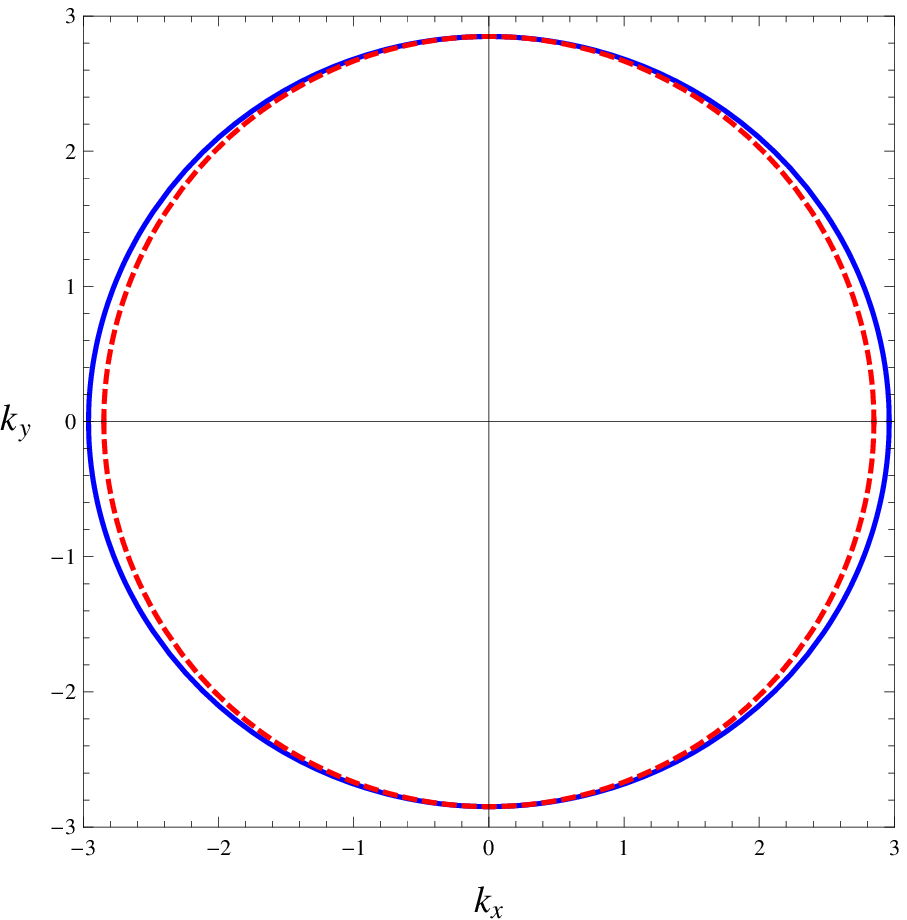}
\includegraphics[width=.32\textwidth]{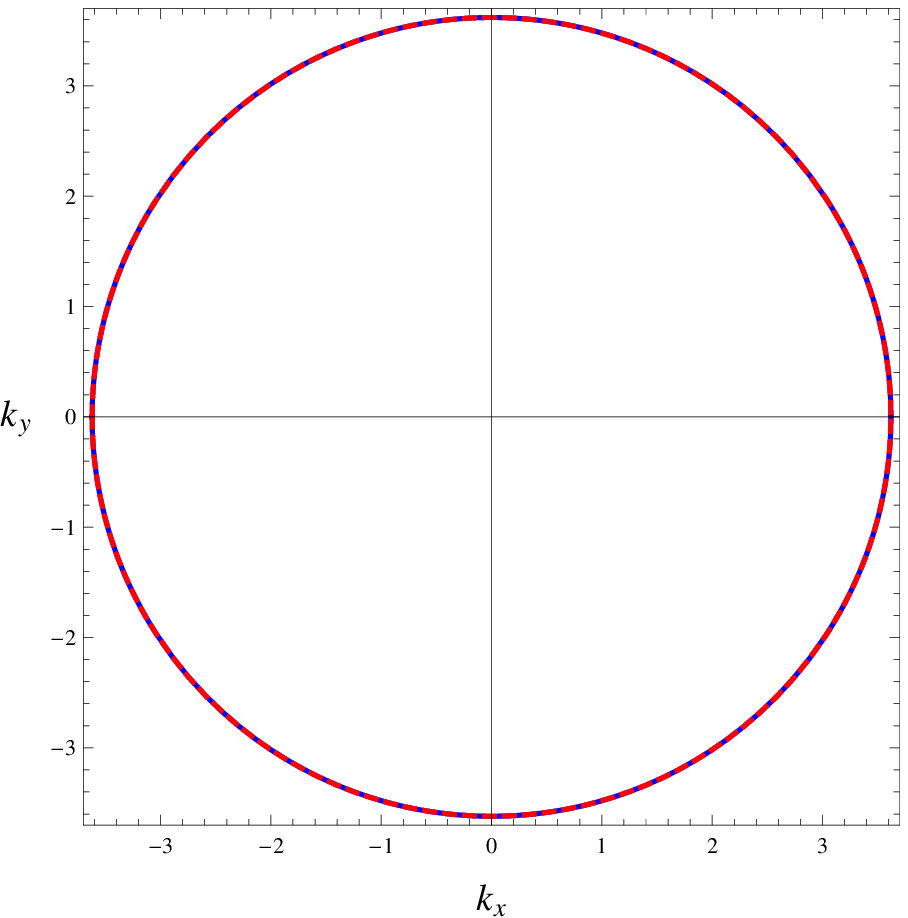}
 \caption{The shape of the Fermi surface for $a^2=1$ (left), $a^2=0$ (right). The red dashed line and blue line represent standard circle and numerical result with different $a^2$, respectively.}
 \label{akxky}
\end{figure}

Turning to the prolate anisotropy $a^2>0$, we expect that the shape of the Fermi surface would be deformed by the anisotropy parameter. Under these circumstances, we work with the stable branch and study the influence of $a^2$ to the Fermi surface.
\begin{table}
\centering
\footnotesize
 \begin{tabular}{|c|c|c|c|c|c|c|c|c|c|c|}
  \hline
  $a^2$   & 0.01        &  0.008       & 0.006       &  0.004       &  0.002        \\ \hline
  $c_{x}$ & 1.82646078  &  1.82973202  & 1.83303540  &  1.83637499  &  1.83975576   \\ \hline
  $c_{y}$ & 1.82599405  &  1.82935522  & 1.83275011  &  1.83618288  &  1.83965868   \\ \hline
  $d$     & 0.00046673  &  0.0003768   & 0.00028529  &  0.00019211  &  0.00009708   \\ \hline
  $f$     & 0.00025554  &  0.000205932 & 0.000155638 &  0.000104614 &  0.0000527679 \\ \hline
 \end{tabular}
\caption{Fermi momentums with different $a^2$ for $a^2>0$}
\label{akfreal}
\end{table}
As the result showed in Table \ref{akfreal}, the Fermi momentum on $k_x$ direction is larger than that on $k_y$ direction and the flattening of the shape of Fermi surface decreases as $a^2$ decreases. In other words, the shape of Fermi surface is a prolate sphere when $a^2>0$\footnote{As a toy model, we left the ``oblate" anisotropy $a^2<0$ case in Appendix \ref{oblate}.}. To have an intuitive picture about the effect of anisotropy parameter $a^2$ on the shape of the Fermi surface, we plot the Fermi surface with larger anisotropy in Fig.\ref{akxky}. Note that we set $r_H=1$, $q=1$ and $q_f=5$. The temperature in this case will become bigger than the above discussion. We can find that the dual boundary field theory generates a ``prolate" Fermi surface when $a^2>0$.

\begin{figure}
\centering
\includegraphics[width=.39\textwidth]{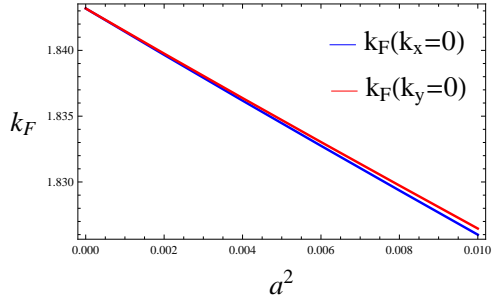}
 \caption{Fermi momentum $k_{F}(k_y=0)$ (red) and $k_{F}(k_x=0)$ (blue) varies with $a^2>0$.}
 \label{kfa2}
\end{figure}

So far, we have examined  the effect of the anisotropy parameter on the shape of the Fermi surface. Choosing the momentum along the anisotropic direction, we reveal that an anisotropic geometry indeed results in an anisotropic Fermi surface. It would be interesting to give an expression for its influence.   In Fig. \ref{kfa2}, we find the Fermi momentum   decreases as $a^2$ increases whether for positive $a^2$. We can obtain the relation between $k_F$ and $a^2$, which is almost linear from Fig.\ref{kfa2}. Then, by fitting the data of $k_{F}$ and $a^2$, we have
\begin{eqnarray}\label{kfafit}
k_{F}(k_y=0)&\sim&1.84318 - 1.72065 a^2 + \mathcal{O}(a^4),\nonumber\\
k_{F}(k_x=0)&\sim&1.84318 - 1.76953 a^2 + \mathcal{O}(a^4).
\end{eqnarray}
From the above relations, we find that the influence of the anisotropy of the background to Fermi momentum on $k_x$ is bigger than on $k_y$ direction. The ``prolate" solution correspond to ``prolate" Fermi surface.
In a word, the anisotropy of the background effect the shape of the Fermi surface.

We emphasize that in the above discussions, the rotational symmetry in the $y-z$ plane is unbroken, since the axion $\chi$ varies linearly in the $x-$direction. Thus, the Fermi surface is isotropic in the $k_y-k_z$ plane.
We also find that smaller anisotropy $a^2$ yields bigger Fermi momentum.

\subsection{The scaling behavior}
\begin{figure}
\centering
\includegraphics[width=.30\textwidth]{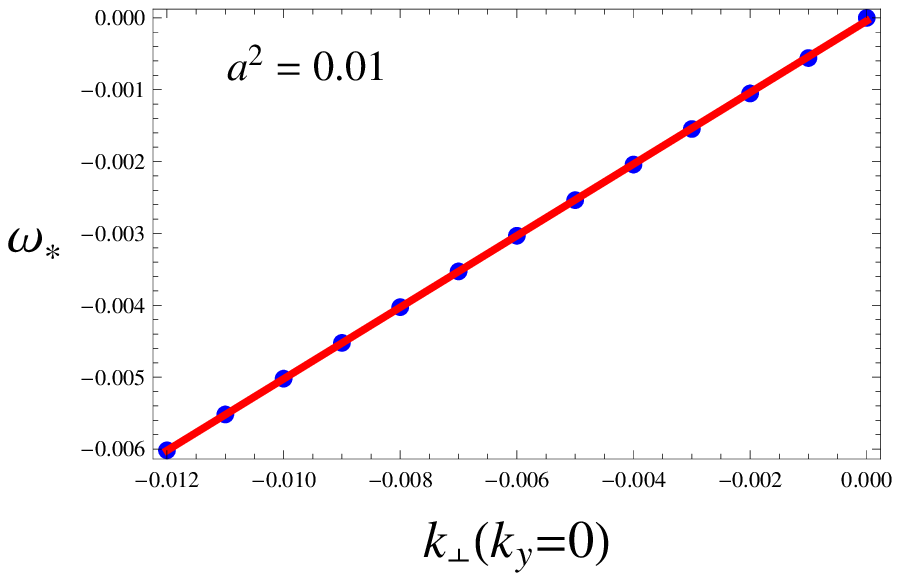}\hspace{0.2cm}
\includegraphics[width=.30\textwidth]{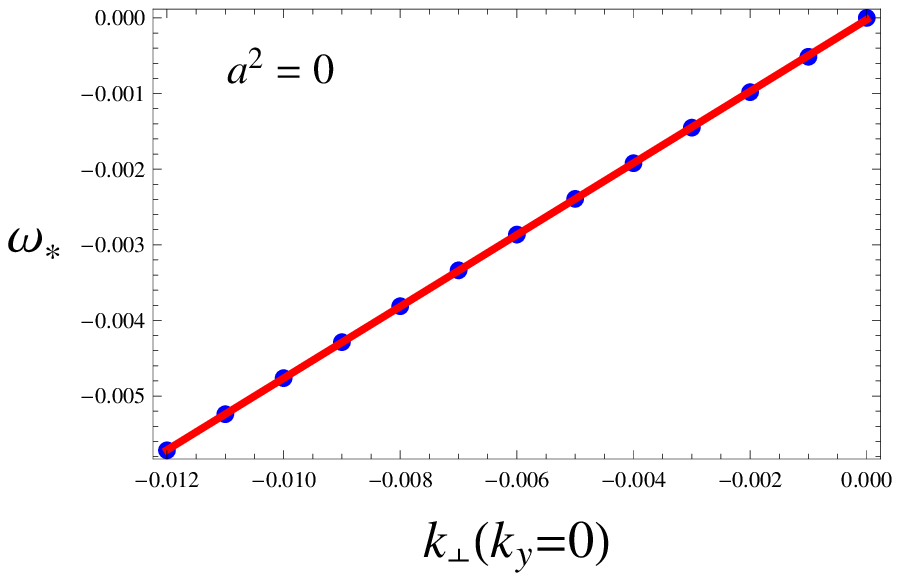}\vspace{0.21cm}\\
\includegraphics[width=.30\textwidth]{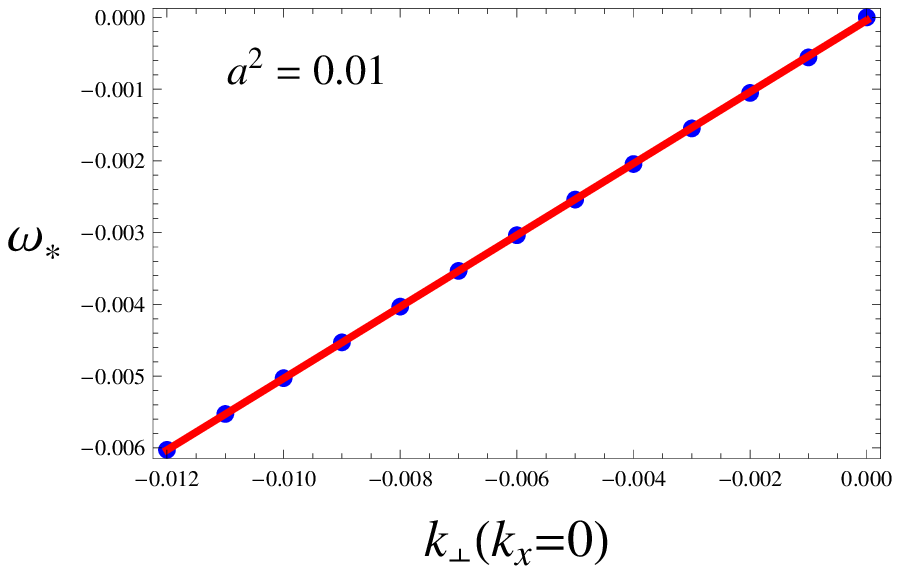}\hspace{0.2cm}
\includegraphics[width=.30\textwidth]{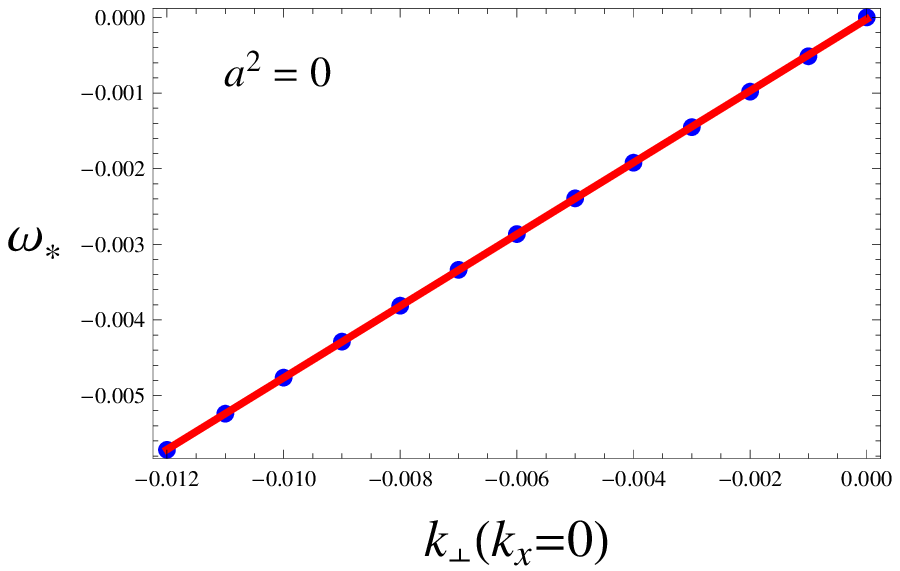}

\caption{The dispersion relation for $a^2=0.01$(left) and $a^2=0$(right).}
\label{drfig}
\end{figure}
\begin{figure}
\centering
\includegraphics[width=.30\textwidth]{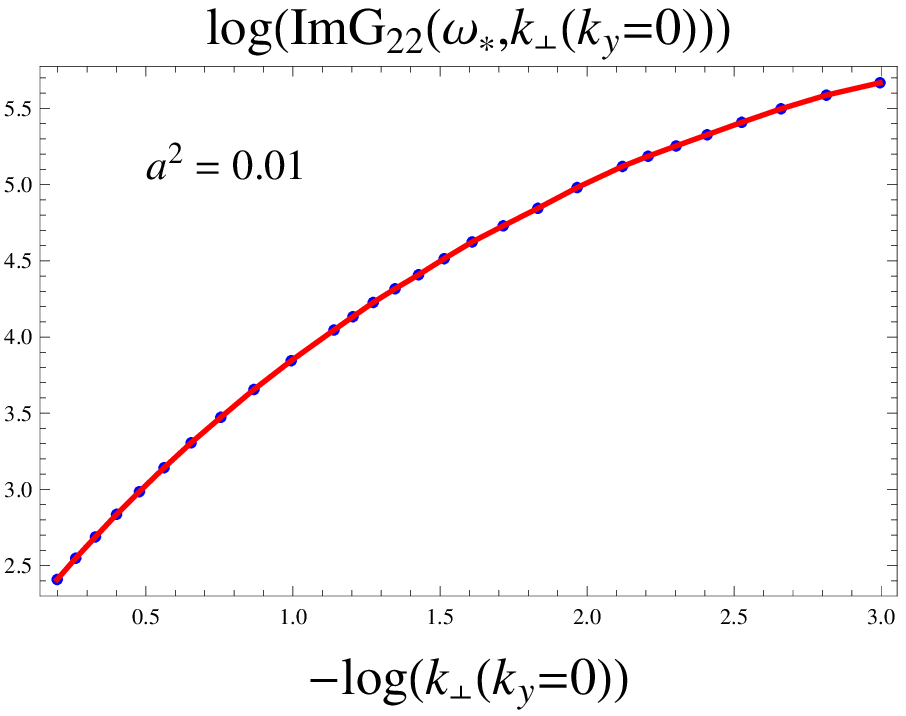}\hspace{0.2cm}
\includegraphics[width=.30\textwidth]{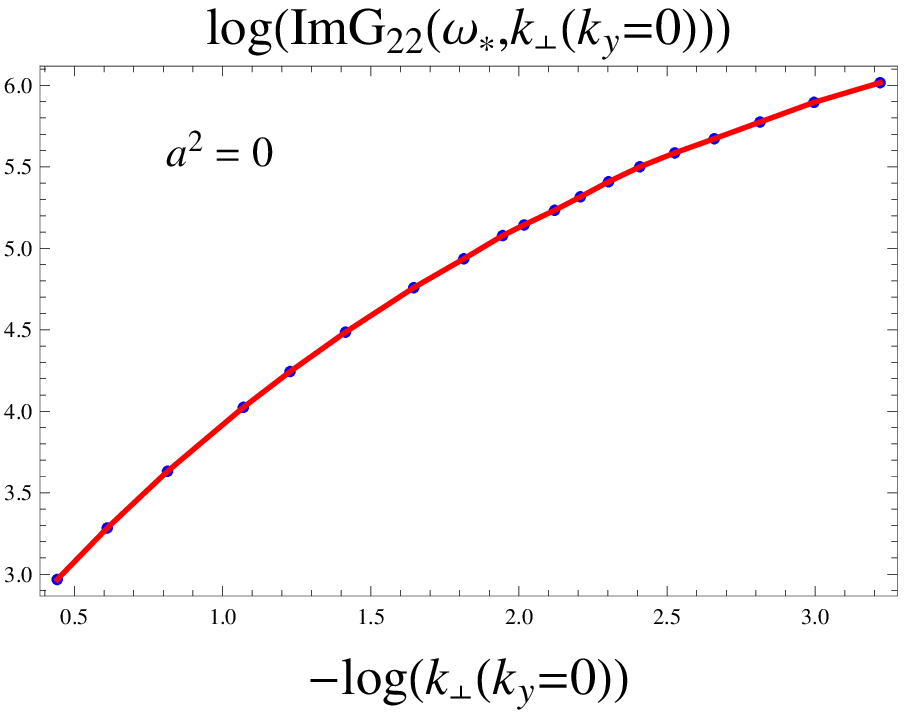}\vspace{0.21cm}\\
\includegraphics[width=.30\textwidth]{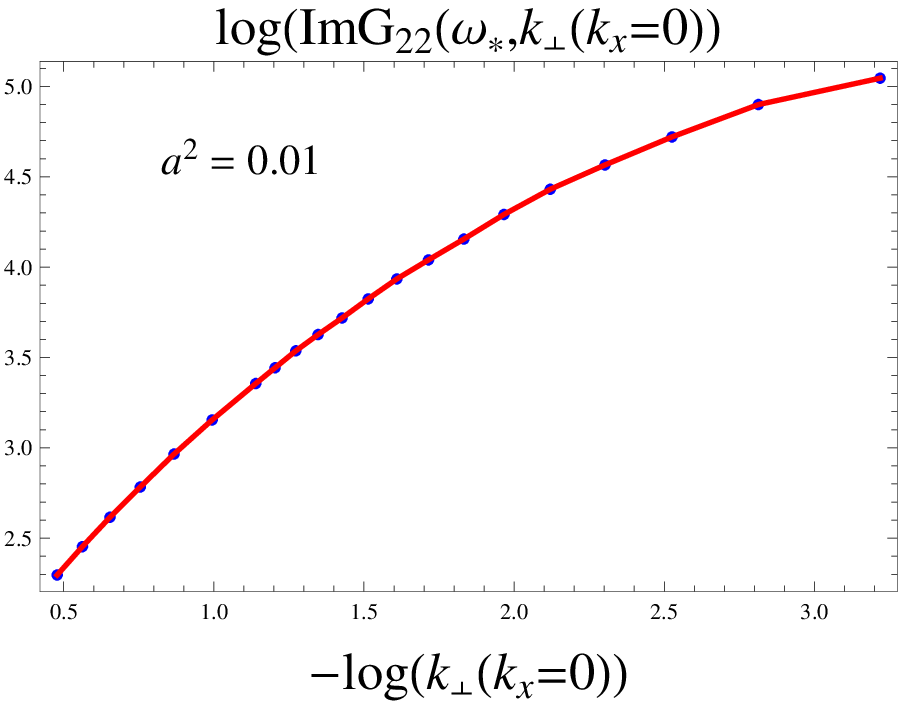}\hspace{0.2cm}
\includegraphics[width=.30\textwidth]{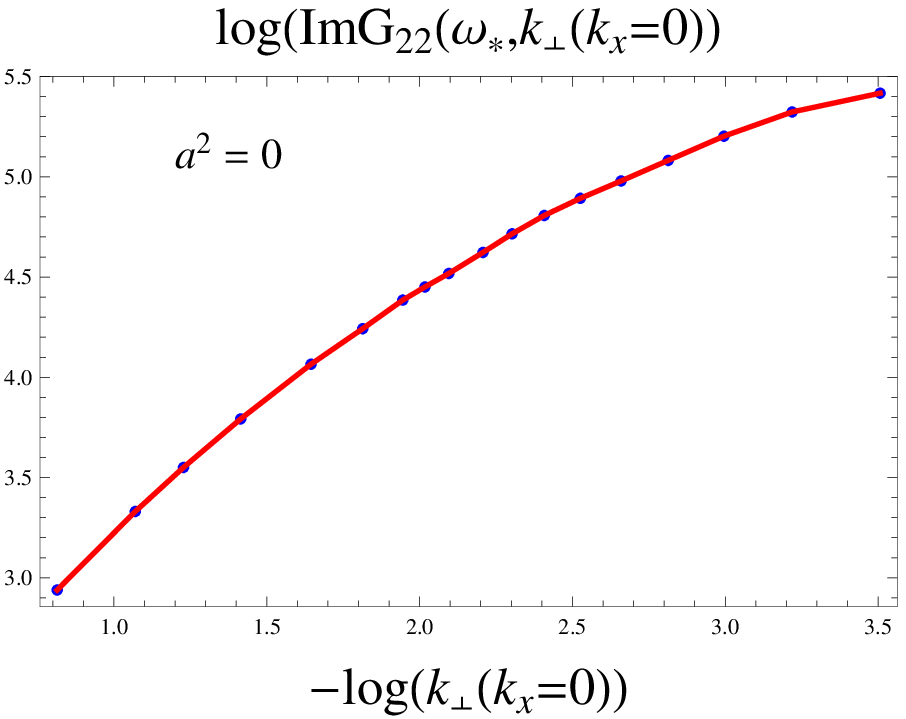}

\caption{The scaling relation of the logarithm of the height of $ImG_{22}(\omega_{\ast}(k_{\perp}),k_{\perp})$ at the maximum for $a^2=0.01$(left) and $a^2=0$(right).}
\label{scalingfig}
\end{figure}

In the following, we try to ascertain the type of the dual system. Thus, we must acquire two scaling behaviors at $k_{\perp}=k-k_{F}\rightarrow0_{-}$. For convenience, we analyze the scaling behaviors with two cases ($k_x=0$ or $k_y=0$).  Denoting the location of the maximum of the quasiparticle-like peak as $\omega_{\ast} (k_{\perp})$, we can write the dispersion relation as
\begin{eqnarray}\label{dr}
\omega_{\ast}(k_{\perp})&\sim& k_{\perp}^{\alpha}.
\end{eqnarray}
As to the case $a^2=0.01$, we find the dispersion relation for this anisotropic background is almost linear (i.e., $\alpha\approx1$)(see the left plot of Fig. \ref{drfig}). To make sure whether the dual liquid is that of the Landau Fermi liquid type, we must check another scaling relation which is the scaling relation of the height of $ImG_{22}$ scales as $k_{\perp}$
\begin{eqnarray}\label{Gkp}
ImG_{22}(\omega_{\ast}(k_{\perp}),k_{\perp})&\sim& k_{\perp}^{-\beta}.
\end{eqnarray}
To be more clearly, we take the logarithm of the both sides of the equation (\ref{Gkp}). As demonstrated in the left plot of Fig.\ref{scalingfig}, we find $\beta\neq1$.
 We can see that $a^2$ cannot change basic type of two scaling behaviors both for $k_x$ and $k_y$ from Fig.\ref{drfig} and Fig.\ref{scalingfig}.

As pointed out in \cite{Senthil1,Senthil2}, the exponent of scaling behavior obeys $\alpha=\beta=1$ for Landau Fermi liquid. We check all the $a^2>0$ and $a^2=0$ cases, which the scaling behavior $\alpha\approx1$ and $\beta\neq1$ is still valid.
So in this respect, the dual liquid does not behave as a Landau Fermi liquid one, which means the type of dual liquid is non-Fermi liquid for this anisotropy background geometry.
For non-Fermi liquids, the transport properties of the systems should be different from that of the Landau Fermi liquid. As an application of the numerical results, one may expect the fermion properties obtained here can be used for understanding the transport properties of the dual field theory.

\subsection{Fermi velocity and the diffusion bound}
As a side note,  we would like to check Hartnoll's conjecture on the diffusion constant. In the following, we focus on the prolate anisotropy $a^2>0$ only. The diffusion constants $D_{+}$ and $D_{-}$ are related to the transport coefficients through the Einstein relations,
\begin{eqnarray}
&&D_{+} D_{-}=\frac{\sigma}{\chi} \frac{\kappa}{c_{\rho}},\label{Dpm1}\\
&&D_{+}+ D_{-}=\frac{\sigma}{\chi}+ \frac{\kappa}{c_{\rho}}+\frac{T(\zeta \sigma-\chi \alpha)}{c_{\rho} \kappa^2 \sigma },\label{Dpm2}
\end{eqnarray}
where $\sigma$, $\alpha$ and $\kappa$ are the electric, the thermoelectric and the thermal conductivities, respectively; $c_{\rho}=T (\partial s/\partial T)_{\rho}$ is the specific heat at fixed charge density, $\chi=(\partial \rho/\partial \mu)_{T}$ is the compressibility and  $\zeta$ is
the thermoelectric susceptibility. In many materials of interest the  thermoelectric related quantities $\alpha$ and $\zeta$ are expected or measured to be small compared to the other terms in (\ref{Dpm2}). In this approximation, we can drop the last term in (\ref{Dpm2}) and conclude that
\begin{equation}
D_{+}=\frac{\sigma}{\chi},~~~~
 D_{-}=\frac{\kappa}{c_{\rho}}.
\end{equation}
Hartnoll's conjecture on the diffusion bound can then be recast as
\begin{equation} \label{bound32}
\frac{\sigma}{\chi},\frac{\kappa}{c_{\rho}}\gtrsim \frac{\hbar v^2_{F}}{k_{B}T}.
\end{equation}
The DC electric conductivity was obtained in  \cite{glns} as
\begin{equation}
\sigma=r_{H}e^{\frac{\phi(r_H)}{4}}+12 r^3_H e^{-\frac{3\phi(r_H)}{4}}\frac{q^2}{a^2}.
\end{equation}
For the given parameters $q=1.4$, $r_H=1$, $a=0.1$ and  the natural unit $\hbar=k_{B}=c=1$, we have $\frac{\sigma}{\chi}=183.89$.
On the other hand, from the dispersion relation (\ref{dr}), we can determine the Fermi velocity $v_F\simeq 0.459756$ numerically. It is easy to obtain $\frac{v^2_{F}}{T}=21.59$. For the cases with $a=0.01$, $a=0.006$ and $a=0.08$, the values of ${v^2_{F}}/{T}$ change slowly and are much  smaller than ${\sigma}/{\chi}$. After a complete analysis, we then conclude that
\begin{equation}
\frac{\sigma}{\chi} > \frac{\hbar v^2_{F}}{k_{B}T}.
\end{equation}
A Similar result can also be obtained for ${\kappa}/{c_{\rho}}$ by counting the intrinsic contribution only \cite{hartnoll2015}. Note that the parameters chosen correspond  to the large horizon (i.e. stable) branch of
the black brane solution.  As to the small horizon  branch of the black brane solution, the specific heat becomes negative. We expect that the bound given in (\ref{bound32}) will be violated.
We defer a thorough investigation on the relation between the shear viscosity bound and the intrinsic diffusivity bound in our anisotropic setup in a future work.

\section{Discussion and Conclusion}\label{sec5}
We have investigated the anisotropic fermions system at low temperature by using a five dimensional charged and anisotropic geometry. The properties of the Fermi surface can be summarized as follows:
\begin{itemize}
  \item The Fermi momentum decreases as the anisotropy parameter $a^2$ increases for all cases.
  \item As the anisotropy parameter $a^2$ increases, the dispersion relation is much closer to the Fermi liquid type.
  \item For the case the momentum along the anisotropic direction, the ``prolate anisotropy" results in ``prolate" Fermi surfaces. On the other hand, the Fermi surface still remains isotropic if the momentum of the Dirac wave function is chosen to be perpendicular to the anisotropic direction.
  \item The scaling behavior has been studied in this background and the dual system is a non-Fermi type liquid. Therefore, the quasiparticle description loses its validity.
\end{itemize}
In general, we obtain an anisotropic non-Fermi liquid from holographic system. The results obtained here can be regarded as direct evidence supporting the conclusion obtained in \cite{glns} that dual transport properties of
the anisotropic black brane is the same as that of the bad metals. As a side note, we verified that our system obeys the recently conjectured universal bound for thermoelectric diffusion constants \cite{hartnoll2015} for the stable branch of the black brane solutions.

For real materials, it always has the lattice structure that corresponds to the Brillouin zone, and the first Brillouin zone is important. In this paper, the axion field does not result in a periodic deformation of the boundary conformal field theory. So the Fermi surface observed here is different from the anisotropic Fermi surface obtained by using a neutral scalar field with periodic boundary conditions  \cite{lattice}.  The Fermi surface which we discussed in this work should locate in  the first Brillouin zone.

\section*{Acknowledgements}
The authors would like to thank Long Cheng and Chao Niu  for their helpful discussions.  This work was partly supported by NSFC,
China (No. 11375110, 11305018, 11275208). X.H.G. would like to thank APCTP for hospitality during the focus program ``Aspects of Holography". J.P.W. was also supported by the Program for Liaoning Excellent Talents in University (No. LJQ2014123).

\appendix
\section{Fermi surface of  oblate  anisotropy  }\label{oblate}
In this section, we attempt to discuss the Fermi surface which is dual to the ``oblate" black brane solution. One may notice that the anisotropy parameter $a$ acts as an isotropy-breaking external source that forces the system into an anisotropic equilibrium state \cite{mateos1}. The $\theta$-parameter is dual to the type IIB axion $\chi$ with the form $\chi = a x$. From the bulk point of view, the  parameter $a$ only plays the role of anisotropy and does not add new degrees of freedom to the SYM theory \cite{mateos1}. However, on the dual quantum field theory side, imaginary $a$ looks like   a  nonunitary deformation and could lead to a negative field coupling. In this sense, the oblate black brane solution with $a^2<0$ could give unphysical results. In this appendix section, we simply take the oblate anisotropy as a toy model and discuss the holographic Fermi surface only for completeness of mathematical computation, which might provide some interesting hints on the dual (non-)Fermi liquid behavior as we see below.

\subsection{Thermodynamic properties of  oblate black brane}
\begin{figure}
\centering
\includegraphics[width=.39\textwidth]{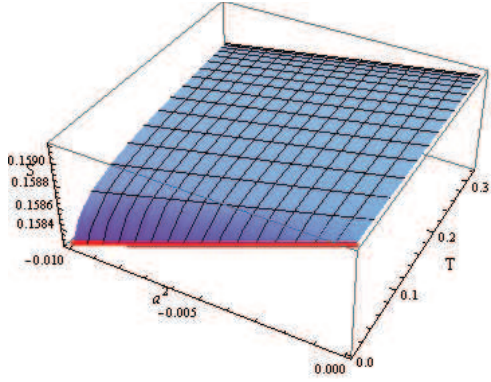}
\caption{The plot of the relation between temperature and entropy for a variational $a^2<0$. }
\label{aTS1}
\end{figure}

 As can be seen from Fig.\ref{aTS1}, the thermodynamics of the oblate black brane is qualitatively the same as the planar black brane \cite{axionbackground1,axionbackground2}. That is to say, there is only one stable branch of black brane solution and the thermodynamics is dominated by this phase for all temperatures.

\subsection{Fermi surface of oblate anisotropy  }

As to the oblate anisotropy, the spectral function can still be solved numerically although the anisotropy parameter becomes imaginary. We find $c_x$ is smaller than $c_y$, and the difference between the $c_x$ and $c_y$ increases as $a^2$ decreases (Table \ref{akfimg}).  The flattening factor of Fermi surface is minus, so the Fermi surface is an ``oblate"-like one. We also find the Fermi momentum decreases as $a^2$ increases for $a^2<0$.
\begin{table}
\centering
\footnotesize
 \begin{tabular}{|c|c|c|c|c|c|c|c|c|c|c|c|}
  \hline
  $a^2$   & -0.002        & -0.004        & -0.006        & -0.008        & -0.01          \\ \hline
  $c_{x}$ &  1.84666749   &  1.85021698   &  1.85384682   &  1.85757781   &  1.86144205    \\ \hline
  $c_{y}$ &  1.84676681   &  1.85041818   &  1.85415283   &  1.85799214   &  1.86196906    \\ \hline
  $d$     & -0.00009932   & -0.0002012    & -0.00030601   & -0.00041433   & -0.00052701    \\ \hline
  $f$     & -0.0000537834 & -0.000108744  & -0.000165068  & -0.000223049  & -0.000283119    \\ \hline
 \end{tabular}
\caption{Fermi momentums with different $a^2$ for $a^2<0$}
\label{akfimg}
\end{table}

\begin{figure}
\centering
\includegraphics[width=.30\textwidth]{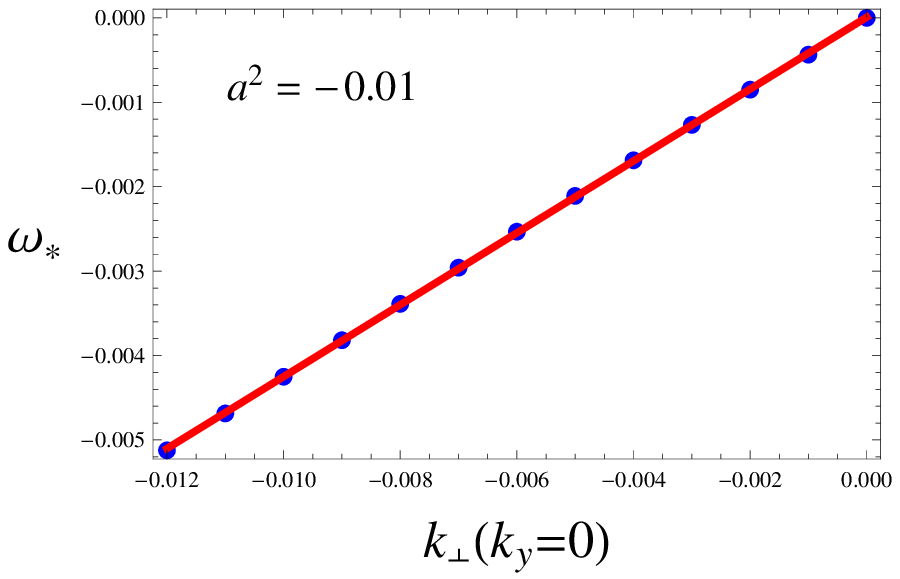}\hspace{0.21cm}
\includegraphics[width=.26\textwidth]{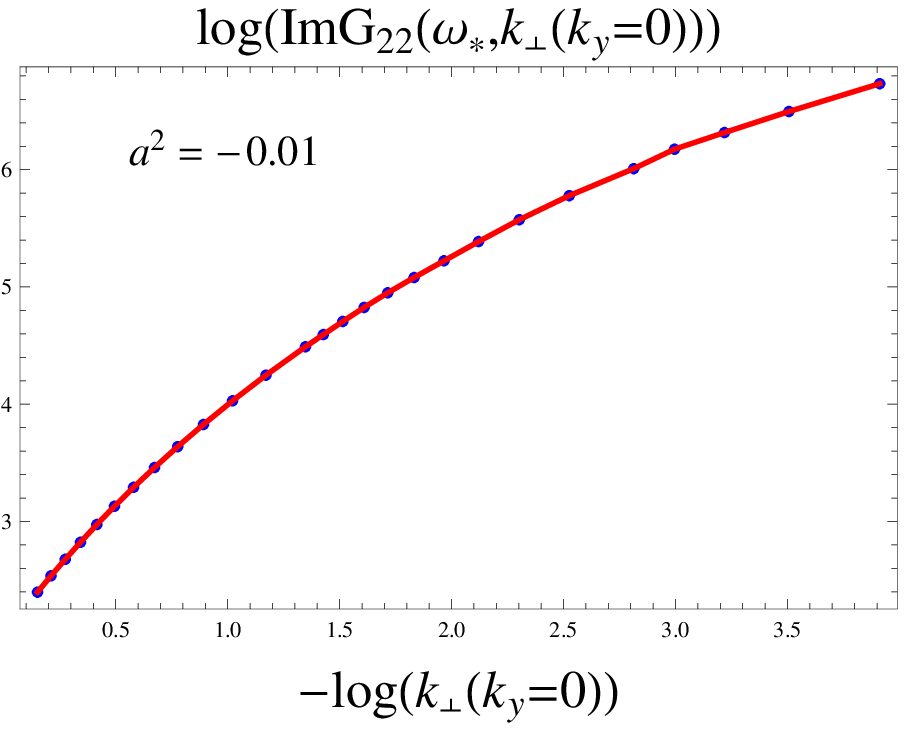}\vspace{0.21cm}\\
\includegraphics[width=.30\textwidth]{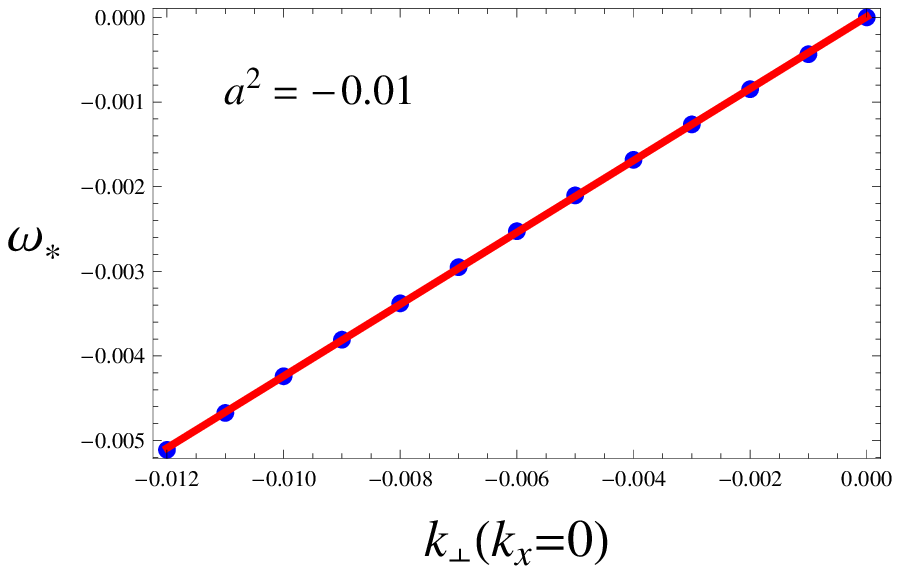}\hspace{0.21cm}
\includegraphics[width=.26\textwidth]{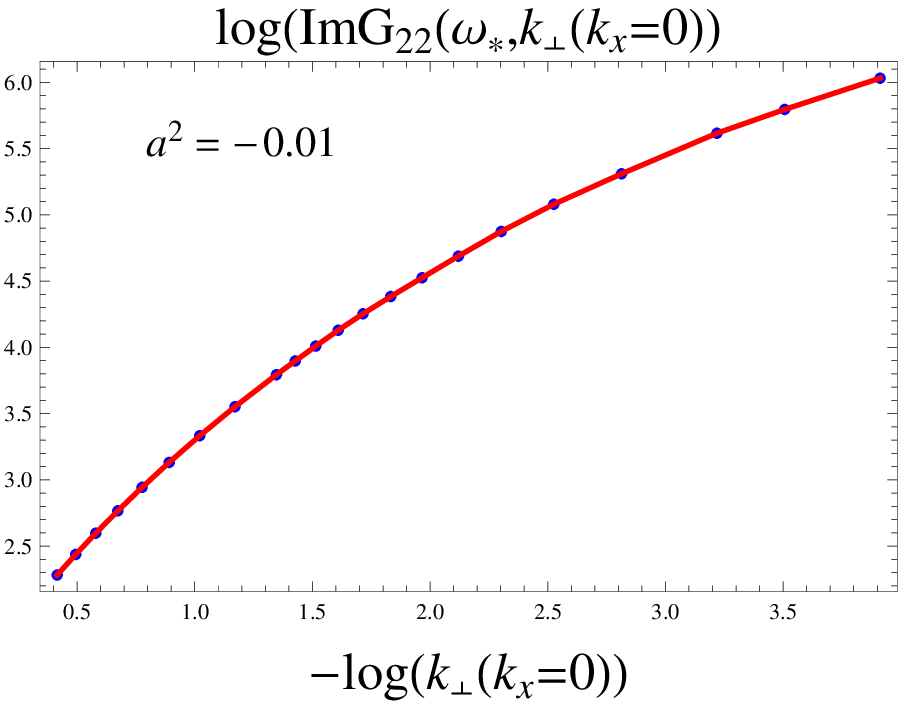}
\caption{The dispersion relation and the scaling relation of the logarithm of the height of $ImG_{22}(\omega_{\ast}(k_{\perp}),k_{\perp})$ at the maximum
for $a^2=-0.01$(right).}
\label{scalingfigimg}
\end{figure}
We also check two scaling behaviors Eq.\eqref{dr} and Eq.\eqref{Gkp} of the Fermi surface for oblate case. As shown in Fig.\ref{scalingfigimg}, we find that two scaling behaviors obey $\alpha\approx1$ and $\beta\neq1$ for both directions, which is similar as the situation of ``prolate" one.
By fitting the data, we obtain the values of $\alpha$ which were summarized in Table \ref{drtable}\footnote{Here, we only calculate the $a^2<-0.006$ cases. Because of the accuracy of our numerical calculation, the fitting method is out of work when the exponent $\alpha$ is close to 1. But we find the same change tendency of the influence of $a^2$ on dispersion relation both for $a^2<0$ and $a^2>0$ with smaller $q_f$ (for example: $q_f=0.5$, $q_f=0.25$). }. We find that the exponent of the dispersion relation $\alpha$ decreases as $a^2$ increases.

\begin{table}[h]
\centering
\footnotesize
 \begin{tabular}{|c|c|c|c|c|c|c|c|c|c|c|c|}
  \hline
  $a^2$ & -0.006 & -0.008 & -0.001      \\ \hline
  $\alpha(k_x)$ &  1.00045 & 1.00490 & 1.01100        \\ \hline
  $\alpha(k_y)$ &  1.00035 & 1.00476 & 1.01091       \\ \hline
 \end{tabular}
\caption{Dispersion relation exponents with different $a^2$ for $q=1$.}
\label{drtable}
\end{table}
Finally, we find the ``oblate" solution leads to ``oblate" Fermi surfaces. The dual liquid is also non-Fermi liquid for such oblate anisotropy.


\begin{thebibliography}{99}
\bibitem{weyl1}F. D. M. Haldane, ``Berry Curvature on the Fermi Surface: Anomalous Hall Effect as a Topological Fermi-Liquid Property", Phys. Rev. Lett. 93, 206602 (2004).
\bibitem{weyl2}Ki-Seok Kim, Heon-Jung Kim, M. Sasaki, ``Anomalous transport phenomena in Weyl metal beyond the Drude model for Landau's Fermi liquids",
arXiv:1407.3056.

\bibitem{adscft}
J. M. Maldacena, ``The Large N Limit of Superconformal Field Theories and Supergravity", Adv. Theor. Math. Phys.  2  231 (1998)
{[arXiv:hep-th/9711200]}.

\bibitem{gkp}
S. S. Gubser, I. R. Klebanov and A. M. Polyakov, ``Gauge Theory Correlators from Non-Critical String Theory", Phys. Lett.
{\bf B 428} 105 (1998) {[arXiv:hep-th/9802109]}.

\bibitem{w}
E. Witten, ``Anti De Sitter Space And Holography", Adv. Theor. Math. Phys. {\bf 2} 253 (1998)
{[arXiv:hep-th/9802150]}.

\bibitem{f1}S. S. Lee, ``A Non-Fermi Liquid from a Charged Black Hole; A Critical Fermi Ball,"
Phys. Rev. {\bf D 79} 086006 (2009)
[arXiv:0809.3402 [hep-th]].
\bibitem{f2} H. Liu, J. McGreevy and D. Vegh, ``Non-Fermi liquids from holography,"
 Phys. Rev. {\bf D 83}   065029 (2011)
[arXiv:0903.2477 [hep-th]].
\bibitem{f3} T. Faulkner, H. Liu, J. McGreevy and D. Vegh, ``Emergent quantum criticality, Fermi surfaces, and AdS2," Phys. Rev. {\bf D 83} 125002 (2011)
[arXiv:0907.2694 [hep-th]].
\bibitem{f4} M. Cubrovic, J. Zaanen and K. Schalm, ``String Theory, Quantum Phase Transitions and the Emergent Fermi-Liquid," Science {\bf 325 } 439 (2009)
[arXiv:0904.1993 [hep-th]].
\bibitem{IL}N. Iqbal and H. Liu, ``Real-time response in AdS/CFT with application to spinors,"
Fortsch. Phys. {\bf 57} 367 (2009),
[arXiv:0903.2596 [hep-th]].
\bibitem{JPW1}J. P. Wu, ``Holographic fermions in charged Gauss-Bonnet black hole," JHEP 07 (2011) 106,
[arXiv:1103.3982 [hep-th]].
\bibitem{JPW2}J. P. Wu, ``Some properties of the holographic fermions in an extremal charged dilatonic black hole
," Phys. Rev. {\bf D 84 } 064008(2011)
[arXiv:1108.6134 [hep-th]].

\bibitem{FLQ1}L. Q. Fang, X. H. Ge, X. M. Kuang,
``Holographic fermions in charged Lifshitz theory", Phys. Rev. {\bf D 86} 105037 (2012).

\bibitem{FLQ2} L. Q. Fang, X. H. Ge, X. M. Kuang,
``Holographic fermions with running chemical potential and dipole coupling", Nucl. Phys. {\bf B 877} 807 (2013).

\bibitem{XMK}X. M. Kuang, E. Papantonopoulos, B. Wang, J. P. Wu, ``Formation of Fermi surfaces and the appearance of liquid phases in holographic theories with hyperscaling violation", [arXiv:1409.2945].

\bibitem{KWP1} X. M. Kuang, B. Wang, J. P. Wu, ``Dipole Coupling Effect of Holographic Fermion in the Background of Charged Gauss-Bonnet AdS Black Hole",  JHEP 07  125 (2012).

\bibitem{KWP2} X. M. Kuang, B. Wang, J. P. Wu, ``Dynamical gap from holography in the charged dilaton black hole",  Class. Quantum Grav. {\bf 30}  145011 (2013).

\bibitem{axionbackground1}
L. Cheng, X. H. Ge and S. J. Sin, ``Anisotropic plasma with a chemical potential and scheme-independent instabilities", Phys. Lett. {\bf B 734} 116 (2014) [arXiv:1404.1994[hep-th]].
\bibitem{axionbackground2}
L. Cheng, X. H. Ge and S. J. Sin, ``Anisotropic plasma at finite $U(1)$ chemical potential",  JHEP {\bf 07} 083 (2014)  [arXiv:1404.5027[hep-th]].

\bibitem{mateos1}D. Mateos and D. Trancanelli, ``The anisotropic N = 4 super Yang-Mills plasma and its instabilities", Phys. Rev. Lett. {\bf 107} 101601 (2011)
\bibitem{mateos2}D. Mateos, D. Trancanelli, ``Thermodynamics and instabilities of a strongly coupled anisotropic plasma", JHEP {\bf 1107} 054 (2011).

\bibitem{glns} X. H. Ge, Y. Ling, C. Niu and S. J. Sin, ``Holographic transports and stability in anisotropic linear axion model",  arXiv:1412.8346 [hep-th]


\bibitem{pomeranchuk} M. Edalati, K. W. Lo, P. W. Phillips, ``Pomeranchuk instability in a non-Fermi liquid from holography", Phys. Rev. {\bf D 86} 086003 (2012).

\bibitem{lattice} Y. Ling, C. Niu, J. P. Wu, Z. Y. Xian, H. Zhang, ``Holographic Fermionic Liquid with Lattices", JHEP {\bf 07} 045 (2013)  [arXiv:1304.2128[hep-th]]

\bibitem{hartnoll2015} S. A.Hartnoll, ``Theory of univerisal incoherent metallic transport", Nature Physics  {\bf 11} 54 (2015) [arXiv:1405.3651[cond-mat.str-el]]


\bibitem{Senthil1}T. Senthil, ``Critical fermi surfaces and non-fermi liquid metals", Phys. Rev. B 78, 035103 (2008),
[arXiv:0803.4009 [cond-mat.str-el]].
\bibitem{Senthil2}T. Senthil, ``Theory of a continuous Mott transition in two dimensions", Phys. Rev. B 78, 045109 (2008),
[arXiv:0804.1555 [cond-mat.str-el]].




\end{thebibliography}
\end{document}